\documentclass[useAMS,usenatbib,twocolumn]{mnras}
\usepackage{graphicx}
\usepackage{dcolumn}   
\usepackage{bm}        
\usepackage[dvipsnames]{xcolor}
\usepackage{amssymb, amsmath,framed,mathtools}

\expandafter\ifx\csname package@font\endcsname\relax\else
 \expandafter\expandafter
 \expandafter\usepackage
 \expandafter\expandafter
 \expandafter{\csname package@font\endcsname}
\fi
\def\bq{\begin{equation}}
\def\eq{\end{equation}}
\def\bqy{\begin{eqnarray}}
\def\eqy{\end{eqnarray}}

\def\ga{\gamma}

\def\la{\lambda}

\begin{document}
\title[Photosynthesis around low-mass stars]{Photosynthesis on habitable planets around low-mass stars}

\author[Lingam \& Loeb]{Manasvi Lingam$^{1,2}$\thanks{E-mail:
manasvi.lingam@cfa.harvard.edu} and Abraham Loeb$^{1,2}$\thanks{E-mail:
aloeb@cfa.harvard.edu}\\
$^{1}$Institute for Theory and Computation, Harvard University, Cambridge MA 02138, USA\\
$^{2}$Harvard-Smithsonian Center for Astrophysics, 60 Garden Street, Cambridge, MA 02138, USA}

\pagerange{\pageref{firstpage}--\pageref{lastpage}} \pubyear{2017}

\maketitle

\label{firstpage}

\begin{abstract}
We show that planets around M-dwarfs with $M_\star \lesssim 0.2 M_\odot$ may not receive enough photons in the photosynthetically active range of $400$-$750$ nm to sustain Earth-like biospheres. As a result of the lower biological productivity, it is likely that biotic molecular oxygen would not build up to detectable levels in the atmospheres of habitable planets orbiting low-mass stars, consistent with prior work by \citet{LCP18}. We also estimate the minimum flaring rate for sustaining biospheres with Earth-like productivity and permitting the build-up of atmospheric oxygen, and find that the overwhelming majority of M-dwarfs are unlikely to exceed this threshold.  
\end{abstract}


\section{Introduction} \label{SecIntro}
It is no exaggeration to claim that life on Earth is dominated by photosynthesis as far as its biomass is concerned \citep{BPM18}. Photosynthesis arose relatively early in Earth's evolutionary history, with the emergence of anoxygenic photosynthesis ostensibly serving as the precursor for oxygenic photosynthesis, which involves the splitting of water molecules to produce oxygen \citep{XB02,HB11,FHJ16}. In its simplified form, the net reaction can be expressed as
\begin{equation}\label{PhotOx}
    \mathrm{CO_2} + 2 \mathrm{H_2O} \, \xrightarrow[\text{pigments}]{h \nu}\, \mathrm{CH_2O} + \mathrm{H_2O} + \mathrm{O_2},
\end{equation}
with the presence of H$_2$O on both the left- and right-hand-sides signifying the fact that water serves as both reactant and product. Oxygenic photosynthesis dominates the net primary productivity (entailing the fixation of carbon) on modern-day Earth, and is often regarded as a major evolutionary development because of the dramatic changes that were subsequently engendered \citep{Lane03,Knoll15}.

Given the centrality of photosynthesis for the reasons outlined above, one may naturally ask as to whether other planets are capable of sustaining biospheres whose biomass is similar to that of Earth. One of the key points to note is that the rate of carbon fixation via photosynthesis on Earth is dependent on the flux of photons in the wavelength range of $\sim 400$-$700$ nm, collectively termed photosynthetically active radiation (PAR). In the habitable zone of cool stars (M-dwarfs in particular), it is well-known that the PAR flux is much smaller than the solar value on Earth, implying that biospheres on habitable planets orbiting these stars may be constrained by the availability of photons \citep{Pol79,GW17,LCP18}. Thus, in this paper, we investigate the characteristics of photosynthesis-based biospheres on planets situated in the habitable zone, especially around low-mass stars. 

Earlier studies in this area include \citet{KST07,GW17,RLR18,MB18,LCP18}. Our analysis closely parallels the recent work by \citet{LCP18} and is consistent with their results, but it differs in the following respects. First, we study the maximum biological potential of Earth-like planets and the capacity for accumulating molecular oxygen in the atmosphere not only for M-dwarfs, but also for A-, F-, G- and K-type stars. Second, we analyze the significance of stellar flares in detail and derive an explicit criterion that must be satisfied in order for flares to contribute significantly to the flux of PAR. 

The outline of the paper is as follows. In Section \ref{SecPF}, we compute the PAR flux received by Earth-analogs, assess the contribution from stellar flares, and consider the limits on biological productivity. We explore the ensuing ramifications for the build-up of atmospheric oxygen and the emergence of complex life in Section \ref{SecO2}. Finally, we summarize our primary conclusions in Section \ref{SecConc}.

\section{Photon fluxes at Earth-analogs}\label{SecPF}
We will examine the PAR photon fluxes received by Earth-analogs orbiting other stars and discuss the contribution of stellar flares to the fluxes of PAR.

\subsection{Background UV fluxes}
By Earth-analogs, we refer to planets with the same basic physical parameters as that of the Earth (e.g. radius, effective temperature). Thus, for an Earth-analog orbiting a star with luminosity $L_\star$, its corresponding semi-major axis $a_\star$ is expressible as
\begin{equation}\label{EAdist}
 a_\star = 1\,\mathrm{AU}\,\left(\frac{L_\star}{L_\odot}\right)^{1/2}   
\end{equation}
assuming that the planet's effective temperature and albedo are the same as that of the Earth. Second, for simplicity, we assume that the star is a perfect blackbody with an effective temperature $T_\star$ and with radius $R_\star$, so that its luminosity is given by
\begin{equation}\label{Lumdef}
    L_\star = 4\pi \sigma R_\star^2 \,T_\star^4.
\end{equation}
Lastly, we assume that the wavelength range of photons accessible for photosynthesis is bracketed between $\lambda_\mathrm{min} = 400$ to $\lambda_\mathrm{max} = 750$ nm. In actuality, the theoretical limits of photosynthesis around other stars are not well understood - some studies indicate that photosynthesis at infrared wavelengths may be feasible \citep{WoRa02,KST07} - owing to which we opt to work with the same limits as those on Earth.

\begin{figure}
\includegraphics[width=7.5cm]{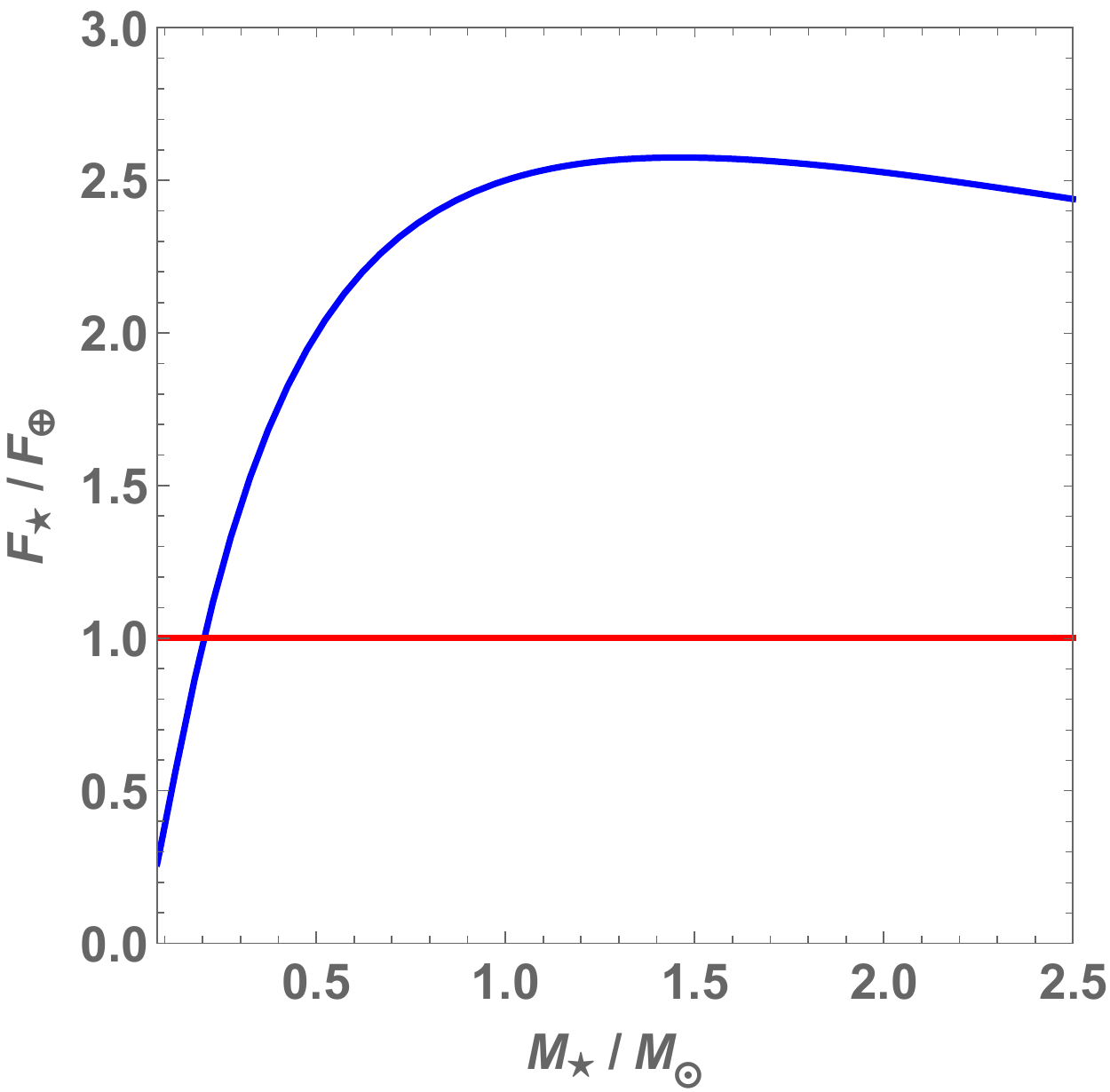} \\
\caption{The photon flux in the wavelength range of $400$-$750$ nm received on the surface of a habitable Earth-analog as a function of the stellar mass ($M_\star$) in units of solar mass ($M_\odot$). The horizontal red line denotes the minimum flux that is required to ensure that the NPP of the biosphere is close to that of the Earth.}
\label{FigFlux}
\end{figure}

Bearing the above caveat in mind, we note that the photon flux $F_\star$ received at the Earth-analog is
\begin{equation}\label{PhoFlu}
  F_\star = \frac{\dot{N}_\star}{4\pi a_\star^2}
\end{equation}
where the photon production rate $\dot{N}_\star$ is defined as
\begin{equation}\label{NstarBB}
\dot{N}_\star = 4 \pi R_\star^2 \times \int_{\lambda_\mathrm{min}}^{\lambda_\mathrm{max}} \frac{2c}{\lambda^4}\left[\exp\left(\frac{h c}{\lambda k_B T_\star}\right)-1\right]^{-1}\,d\lambda,
\end{equation}
and the integrand follows from dividing the spectral radiance $B_\lambda$ of a blackbody by the energy $hc/\lambda$ of a photon. Finally, in order to express $F_\star$ solely in terms of $M_\star$, we use the mass-luminosity and mass-radius relationships, given by $L_\star \propto M_\star^3$ - see Eq. (3.18) of \citet{BV92} and pg. 139 of \citet{SC05} - and $R_\star \propto M_\star^{0.8}$ \citep{JGBL15} respectively, in conjunction with (\ref{Lumdef}). 

The integral (\ref{NstarBB}) is evaluated analytically and substituted into (\ref{PhoFlu}). We are interested in the ratio $F_\star/F_\oplus$, where $F_\oplus$ denotes the critical value of the PAR photon flux incident on the planet that is necessary to sustain a net primary productivity (NPP) equal to that of the Earth. By using the data from \citet{FB98}, it can be shown that $F_\oplus \approx 4 \times 10^{20}$ photons m$^{-2}$ s$^{-1}$. Hence, we have plotted $F_\star/F_\oplus$ as a function of the stellar mass in Figure \ref{FigFlux}. First, it must be noted that $F_\star/F_\oplus \approx 2.5$ for $M_\star = M_\odot$; this arises because the NPP of the Earth is not limited by the availability of photons but by the abundance of nutrients. Second, it can be seen from this figure that Earth-analogs around stars with $M_\star \lesssim 0.21 M_\odot$ do not fulfill the desired requirement of $F_\star/F_\oplus > 1$. Thus, temperate planets orbiting low-mass M-dwarfs have a lower likelihood of sustaining biospheres analogous to the Earth. 

\subsection{Contribution from stellar flares}\label{SSecFla}
However, our analysis has relied on the premise that we can approximately model the stellar spectrum as a blackbody. This assumption is problematic in the case of M-dwarfs since many of them are characterized by regular stellar flares and associated space weather phenomena. Obtaining a general expression for $F_\star$, which includes the contribution from flares, is not feasible since the flaring activity can vary by orders of magnitude even for stars with the same value of $M_\star$ \citep{Dav16}. 

However, a couple of general statements are possible regarding the impact of flares on $F_\star$. First, for quiescent stars such as the Sun, the contribution to $F_\star$ from flares is much lower than the blackbody spectrum. This is because of the fact that the time-averaged flare luminosity (and photon flux) is predicted to be much lower than the bolometric luminosity \citep{Benz17}. Second, for very active M-dwarfs, the contribution from flares might result in the enhancement of $F_\star$ by up to an order of magnitude \citep{MB18}. Therefore, our results are more likely to be valid for comparatively quiescent stars, for which the contribution from flares will not alter the blackbody value of $F_\star$ significantly. 

We can further quantify the requirements for flares to supply enough PAR as follows. A similar approach was delineated in \citet{GZS19} for prebiotic chemistry mediated by UV-C radiation, with the flare modeled approximately as a blackbody at temperature $9000$ K. Let us suppose that the flare energy is denoted by $E_f$ whereas $\dot{\mathcal{N}}_f$ (in units of inverse time) denotes the occurrence rate of flares with this energy. Furthermore, we will suppose that a fraction $\epsilon_\mathrm{PAR}$ of the total flare energy is emitted in the PAR wavelength range. In this case, the photon flux $F_f$ received at the Earth-analog due to flares is given by
\begin{equation}
    F_f \approx \frac{\epsilon_\mathrm{PAR} \dot{\mathcal{N}}_f E_f \bar{\lambda}}{4 \pi h c a_\star^2},
\end{equation}
where we have introduced the extra factor $h c/ \bar{\lambda}$ to convert the energy flux into the photon flux, with $\bar{\lambda}$ denoting the average wavelength of a photon in the PAR range. Since the wavelength spans $\sim 400$-$750$ nm, note that the energy of a photon only varies by a factor of less than $2$. We will use the arithmetic mean of the two energies, thus yielding $h c/ \bar{\lambda} = 3.8 \times 10^{-19}$ J.
In order for an Earth-like NPP to be sustained, the criterion $F_f > F_\oplus$ must be fulfilled. In order to compute $\epsilon_\mathrm{PAR}$, we use the approximation that the flare can be modeled as a blackbody with temperature $9000$ K, which yields $\epsilon_\mathrm{PAR} \approx 0.4$. From this data, we obtain:
\begin{equation}\label{FFReq}
 \dot{\mathcal{N}}_f \gtrsim 9.4 \times 10^3\,\mathrm{day}^{-1}\,\left(\frac{E_f}{10^{27}\,\mathrm{J}}\right)^{-1} \left(\frac{R_\star}{R_\odot}\right)^{2} \left(\frac{T_\star}{T_\odot}\right)^{4},    
\end{equation}
and this can be further simplified using the mass-radius and mass-luminosity scalings introduced earlier to yield
\begin{equation}\label{CFreq}
 \dot{\mathcal{N}}_f \gtrsim 9.4 \times 10^3\,\mathrm{day}^{-1}\,\left(\frac{E_f}{10^{27}\,\mathrm{J}}\right)^{-1} \left(\frac{M_\star}{M_\odot}\right)^{3}.    
\end{equation}
Note that (\ref{FFReq}) has the same functional dependence as Eq. (10) from \citet{GZS19}, except that the prefactor of $9.4 \times 10^3$ in the former must be replaced with $3.4 \times 10^2$. Using the data from the \emph{TESS} mission presented in Figure 10 of \citet{GZS19}, it is easy to verify that $< 1\%$ of all M-dwarf flaring stars satisfy the condition (\ref{FFReq}), implying that flares might not suffice to deliver enough PAR to enable the sustenance of biospheres with Earth-like NPP around the majority of low-mass M-dwarfs. Next, let us apply (\ref{CFreq}) to Proxima Centauri. We find that the cumulative flaring frequency must be $\sim 160$/day for flares with $\gtrsim 10^{25}$ J. Using the cumulative flare frequency distribution for Proxima Centauri deduced from \emph{MOST} observations \citep{DKS16}, we find that this criterion is not fulfilled. Likewise, repeating the same analysis for TRAPPIST-1 using (\ref{CFreq}), the frequency must be $\sim 70$/day for flares with $\gtrsim 10^{25}$ J. Employing the flare frequency distribution formulated from the analysis of the \emph{K2} mission \citep{VKP17}, this condition is not met. Hence, at least insofar as Proxima Centauri and TRAPPIST-1 are concerned, it seems unlikely that their flares deliver enough PAR fluxes to permit the sustenance of Earth-like NPP on planets orbiting these stars. 

Moreover, there are additional caveats that deserve to be highlighted here. Unlike standard photosynthesis that can function continuously during the day, flare-driven photosynthesis would necessitate a very different mode of operation. Hence, in the absence of comprehensive laboratory experiments, it is not clear as to whether intermittent photosynthetic activity is possible. In particular, it is not clear as to whether densely ionizing radiation emitted during solar proton events associated with flares could induce damage and thereby inhibit photosynthesis \citep{DAPP}. Finally, we note that flares give rise to a number of other positive and negative effects. In the former category, they are capable of supplying ultraviolet photons and stellar energetic particles for prebiotic chemistry \citep{BL07,RWS17,MaLi18,RXT18,LDF18}. On the other hand, flares contribute, either directly or indirectly, to a number of detrimental phenomena including atmospheric erosion \citep{LLK07,DLMC,DHL17,DJL18} and ozone depletion \citep{SW10,LL17,LiLo}.

\section{Implications for atmospheric oxygen}\label{SecO2}

\begin{figure}
\includegraphics[width=7.5cm]{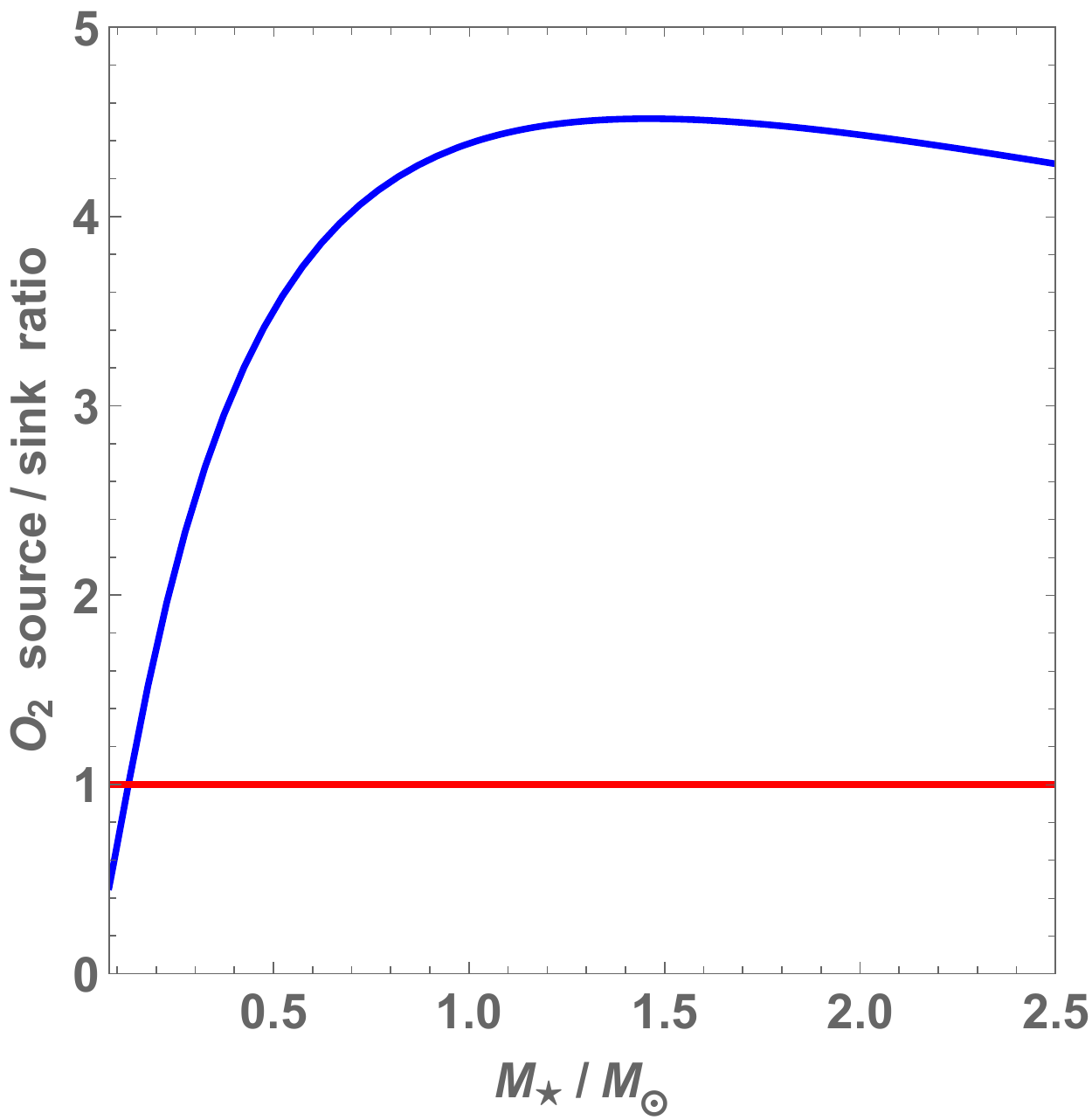} \\
\caption{The maximum possible oxygen source-to-sink ratio ($\Delta_{O_2}$) as a function of the stellar mass ($M_\star$) assuming that the PAR flux constitutes the sole limitation. In the case of Earth-analogs around stars whose values lie below that of the red line (where the source and sink are balanced), oxygen depletion is dominant.}
\label{FigORat}
\end{figure}

Hitherto, we have concentrated on assessing the potential biomass sustainable on Earth-analogs by taking only energetic constraints into account. However, it should be recognized that oxygenic photosynthesis not only entails the fixation of carbon but also the release of oxygen as a product, as seen from (\ref{PhotOx}). Thus, the photon flux also dictates the amount of oxygen released into the atmosphere provided that all the sinks are held fixed. This consideration implicitly presumes that other limitations (e.g. nutrients) do not come into play; we will return to this issue later. 

On Earth, it is known that oxygenic photosynthesis indirectly serves as the source of atmospheric oxygen via the burial of organic matter. A second major source of O$_2$ requires the burial of pyrite (FeS$_2$). There are two major sinks, but only one of them is expected to be dominant in worlds where O$_2$ has not built up to sufficiently high levels. This sink corresponds to O$_2$ consumption by reactions with reducing gases arising from volcanism and submarine weathering \citep{CK17}. \citet{LCP18} recently suggested that a potentially sufficient condition for ensuring oxygen buildup in the atmosphere is that the input rate must exceed the depletion rate. The influx rate of O$_2$ is computed by multiplying the maximum amount of carbon that can be fixed with the efficiency of burial of organic carbon \citep{Holl02}. The former factor in turn is directly proportional to the PAR flux. Following \citet{LCP18}, we hold the depletion rate to be constant, and adopt the value of $\sim 5.7 \times 10^{12}\,\mathrm{mol/yr}$ presented in \citet{CK17}. Further details regarding this procedure can be found in \citet{LiLo18}. In actuality, the oxygen depletion rate is time-dependent and is also likely to vary across different planets depending on their geological properties (e.g. internal heat budget).

In Figure \ref{FigORat}, we have depicted the ratio of the oxygen source and sink fluxes, denoted henceforth by $\Delta_{O_2}$. This quantity only represents the maximum possible value (but not the actual value), since it presupposes that all of the PAR is utilized by photosynthesis in the absence of any other limitations. It is found that $\Delta_{O_2} < 1$ holds true when $M_\star \lesssim 0.13\,M_\odot$. In other words, M-dwarfs with masses below this threshold, which constitute a sizable fraction of all stars including the local examples of Proxima Centauri and TRAPPIST-1, have a low probability of accumulating oxygen in the atmosphere via oxygenic photosynthesis. This result has two important consequences for biosignatures and technosignatures as discussed below.

First, if oxygen levels are not sufficiently high in the atmosphere, it is likely that searches for oxygen - for example, by means of transit spectroscopy \citep{MRA18} - would give rise to ``false negatives'' \citep{ROS17}. In other words, it is conceivable that planets host life but are not detectable by seeking signatures of O$_2$ and O$_3$ because of the simple fact that the concentrations of these gases in the atmosphere would be too low.\footnote{In the case of tidally locked planets around M-dwarfs, it should also be noted that the day-to-nightside contrasts in the concentration of O$_2$ and O$_3$ could be appreciable \citep{CWK18}.} The Earth, for instance, possessed a largely anoxic atmosphere until $\sim 2.4$ Ga with near-modern O$_2$ levels having been achieved only $\lesssim 0.5$ Ga \citep{LRP14,KNo17,CK17,LL18d}. Hence,  Figure \ref{FigORat} can assist in the identification of suitable target planets based on the stars that they orbit \citep{LoL18}.

Second, the rise in atmospheric and oceanic oxygen levels has been proposed to constitute a necessary requirement for the origin of ``complex'' life endowed with high motility \citep{CGZ05,BaSM16}. Although it cannot be naively said that the rise in O$_2$ levels served as the trigger for the diversification of animals \citep{MC14}, there are sufficient grounds for concluding that the former was responsible at least in part for the latter \citep{Kn17}. Hence, if oxygen is truly a prerequisite for complex life, it will also be required for technological intelligence to evolve. In turn, the emergence of the latter opens up the possibility of finding life via technosignatures such as electromagnetic signals and artifacts \citep{Man18}. 

Hitherto, we have not discussed the role of stellar flares in delivering sufficient PAR fluxes to enable the build-up of atmospheric O$_2$ indirectly via oxygenic photosynthesis. The procedure for calculating the desired occurrence rate is identical to that described in Sec. \ref{SSecFla}. The only difference is that the critical threshold that must be exceeded for permitting biotic O$_2$ accumulation is $\sim 0.57\,F_\oplus$. Thus, we find that the analog of (\ref{CFreq}) for the flaring rate is given by
\begin{equation}\label{CFreqO2}
 \dot{\mathcal{N}}_f \gtrsim 5.4 \times 10^3\,\mathrm{day}^{-1}\,\left(\frac{E_f}{10^{27}\,\mathrm{J}}\right)^{-1} \left(\frac{M_\star}{M_\odot}\right)^{3}.    
\end{equation}
As noted in Sec. \ref{SSecFla}, the overwhelming majority of M-dwarf flaring stars do not meet this condition. Hence, it appears unlikely that flares are capable of providing sufficient PAR fluxes to permit the build-up of atmospheric O$_2$ around M-dwarf exoplanets.

In closing, there is an important point that must be mentioned here. Although our model predicts that the rise in O$_2$ levels may be suppressed or delayed on planets around low-mass M-dwarfs, this refers to \emph{biotic} oxygen. In contrast, there are plenty of abiotic mechanisms for the accumulation of O$_2$ in atmospheres of M-dwarf exoplanets, the most common of which involves the photolysis of water and subsequent escape of hydrogen to space \citep{LiLo}. Thus, even if O$_2$ is detected, it is important to apply suitable diagnostics to distinguish between abiotically and biotically produced oxygen \citep{MRA18}. While the buildup of O$_2$ through abiotic channels may come across as being positive from the standpoint of complex life, it is important to recognize that there are potential downsides as well. For example, it could be accompanied by the loss of water or result in extremely thick atmospheres with pressures of hundreds of bars \citep{LB15}.

\section{Conclusions}\label{SecConc}
We have explored the prospects for sustaining Earth-like biospheres (in terms of biological productivity) via photosynthesis on habitable planets around other stars. We modeled the stellar emission spectrum as a blackbody and computed the photon flux incident at the surface of an Earth-analog in the wavelength range of $400$-$750$ nm. The photon flux received in this range sets limits on the biological potential of these worlds, provided that other desiderata such as nutrients and electron donors are fulfilled.

We found that the following broad conclusions arise. First, quiescent stars with $M_\star \lesssim 0.13 M_\odot$ are more likely to possess habitable planets with anoxic atmospheres due to the fact that the build-up of O$_2$ in the atmosphere via oxygenic photosynthesis is suppressed. This has two important consequences. The absence of sufficiently high oxygen levels might preclude complex multicellular life as well as technological intelligence from arising on such planets. However, even in the absence of oxygenic biosignatures, it would be feasible to detect anoxic biosignatures (e.g., methane) in the atmospheres of worlds where O$_2$ has not built up \citep{KOC18,KGI18}. Interestingly, the above criterion is satisfied by both TRAPPIST-1 and Proxima Centauri, thereby potentially indicating that planets in the habitable zone of these stars may not accumulate biotic oxygen in the atmosphere.

Second, stars with $0.13 M_\odot \lesssim M_\star \lesssim 0.21 M_\odot$ are expected to possess habitable planets where the build-up of atmospheric oxygen is feasible, but the associated timescale for reaching modern Earth levels could be much longer (possibly exceeding the age of the Universe) and the overall NPP is predicted to be smaller than that of the Earth. Third, habitable planets around stars with $M_\star \gtrsim 0.21\,M_\odot$ are theoretically capable of sustaining biospheres with the same productivity as the Earth, provided that photon limitation constitutes the only determining factor. Lastly, we calculate the minimum flaring rates necessary for the sustenance of biospheres with Earth-like productivity and the build-up of atmospheric oxygen via oxygenic photosynthesis as a function of stellar mass and flare energy. We concluded that a potentially negligible fraction ($\lesssim 1\%$) of M-dwarfs are likely to exceed the critical flaring frequency. 

Thus, our analysis suggests, in agreement with previous studies, that the likelihood of finding Earth-like biospheres around low-mass stars is comparatively low. In particular, these conclusions are consistent with those of \citet{LCP18} who expressed their results in terms of a lower bound on stellar irradiation. Here, we have employed mass-luminosity and radius-luminosity relationships to determine a lower limit on stellar mass, since the latter could be more readily useful for astronomers. Replacing our mass-luminosity relation with the broken power-law scalings presented in \citet{SC05} and \citet{LBS16} will result in the aforementioned cutoffs $0.13\,M_\odot$ and $0.21\,M_\odot$ being replaced by $0.21\,M_\odot$ and $0.39\,M_\odot$. Hence, our estimates of $0.13\,M_\odot$ and $0.21\,M_\odot$ may serve as fairly robust lower bounds for permitting the build-up of atmospheric O$_2$ and sustenance of Earth-like biological productivity, respectively. If the build-up of atmospheric oxygen and the evolution of complex life is indeed suppressed on low-mass M-dwarfs, this might provide a partial explanation as to why \emph{Homo sapiens} (i.e., intelligent and conscious observers) find themselves around a Sun-like star in the present epoch instead of orbiting an M-dwarf in the cosmic future \citep{LBS16,HMKW}.

Throughout our treatment, we have implicitly assumed that the properties of extraterrestrial biospheres (e.g. photosynthetic pathways) are akin to that of our planet, and we have not taken planetary factors such as the land-water fraction into account \citep{LiLo18}. Moreover, our analysis accounts for stellar mass, but not for other stellar parameters such as age and rotation rate. Despite these caveats, the advantage of this simple model is that it provides predictions for the detectability of biosignatures and technosignatures that will be testable in the coming decades.

\section*{Acknowledgments}
We thank our reviewer, David Catling, for his helpful comments and suggestions that helped improve the paper. This work was supported in part by the Breakthrough Prize Foundation, as well as by the Faculty of Arts and Sciences and the Institute for Theory and Computation (ITC) at Harvard University.


\label{lastpage}

\end{document}